\documentclass{article}

\usepackage[english]{babel}

\usepackage[letterpaper,top=2cm,bottom=2cm,left=1.3cm,right=1.3cm,marginparwidth=1.5cm]{geometry} %

\usepackage{caption}
\usepackage{subcaption}
\usepackage{bbm}
\usepackage{amsmath, amssymb, mathtools}
\usepackage{graphicx}
\usepackage{amsthm}
\newtheorem*{remark}{Remark}
\usepackage[colorlinks=true, allcolors=blue]{hyperref}
\usepackage{bm}
\newcommand*{\R}{\mathbb{R}}
\newcommand{\1}{\mathbbm{1}}
\usepackage[numbers]{natbib}
 \usepackage{authblk}
\usepackage[utf8]{inputenc}

\title{Modelling opinion dynamics \\under the impact of influencer and media strategies}
\author[a,b]{Luzie Helfmann}
\author[a]{Nata\v sa Djurdjevac Conrad} 
\author[c]{Philipp Lorenz-Spreen}
\author[a,d,1]{Christof Sch\"{u}tte}

\affil[a]{Zuse Institute Berlin}
\affil[b]{Potsdam Institute for Climate Impact Research}
\affil[c]{Center for Adaptive Rationality, Max Planck Institute for Human Development}
\affil[d]{Institute of Mathematics, Freie Universit\"at Berlin}
\affil[1]{Corresponding author: schuette@zib.de}

\date{January 2023}

\begin{document}
\maketitle

\begin{abstract}
Digital communication has made the public discourse considerably more complex, and new actors and strategies have emerged as a result of this seismic shift. Aside from the often-studied interactions among individuals during opinion formation, which have been facilitated on a large scale by social media platforms, the changing role of traditional media and the emerging role of "influencers" are not well understood, and the implications of their engagement strategies arising from the incentive structure of the attention economy even less so. Here we propose a novel opinion dynamics model that accounts for these different roles, namely that media and influencers change their own positions on slower time scales than individuals, while influencers dynamically gain and lose followers. Numerical simulations show the importance of their relative influence in creating qualitatively different opinion formation dynamics: with influencers, fragmented but short-lived clusters emerge, which are then counteracted by more stable media positions. Mean-field approximations by partial differential equations reproduce this dynamic. Based on the mean-field model, we study how strategies of influencers to gain more followers can influence the overall opinion distribution. We show that moving towards extreme positions can be a beneficial strategy for influencers to gain followers. Finally, we demonstrate that optimal control strategies allow other influencers or media to counteract such attempts and prevent further fragmentation of the opinion landscape. Our modelling framework contributes to better understanding the different roles and strategies in the increasingly complex information ecosystem and their impact on public opinion formation.
\end{abstract}
\twocolumn
\section*{Introduction}
In the era of many-to-many communication on social media, polarization and radicalism can increasingly be understood as self-organized phenomena emerging from opinion dynamics on social networks \cite{bak2021stewardship}.  
Despite the global use of those large communication platforms and the resulting interactions, they have not been thoroughly studied as effects of a dynamical system consisting of multiple interacting components, particularly distinguishing their different roles and goals.

More specifically, social media did not replace traditional media, but became an intermediary entity that allows peer-to-peer communication among individuals but also traditional media outlets to disseminate their content and interact with their audience. In principle, they are treated like individuals on those platforms, but have very different internal dynamics (e.g., editorial processes, reputation and economic incentives). As more and more people around the world access traditional news media through social platforms, this dependency becomes increasingly pronounced \cite{newman2021reuters}. At the same time a new role emerged in this system, namely, the \emph{influencer}, who is a private person with many followers on a social media network \cite{bakshy2011everyone}. Although mostly following commercial goals, they do turn to politics \cite{zimmermann2020influencers} and can polarize public discussions \cite{soares2018influencers}, as well as compete with traditional media and with one another for audience \cite{tafesse2022social}. Potentially this increased competition could contribute to the declining trust in traditional media and polarized debates on social media around the world \cite{lorenz2022systematic}. We aim to capture the scenario of opinion dynamics under the impact of influencers and media with a generalized modeling framework designed to distinguish these different roles, but allow them to interact with each other.

So far in most cases, opinion dynamics are mathematically modelled by network- or agent-based models (ABMs) that aim to reproduce how individuals change their opinions based on the feedback of their peers in the respective social environment. The largest group of such models describes the change of opinions based on the (dynamical) interaction network between individuals. The well-known voter model \cite{Liggett2005} in which individuals copy the discrete opinion of a random neighbour may serve as an example. Alternative models use continuous opinion spaces where usually individuals' opinions are drawn to attracting opinions in their social network (in some cases repelled by others). The DeGroot model~\cite{degroot1974reaching} with individuals being drawn to the weighted average opinion of their neighbours and bounded confidence models~\cite{hegselmann2002opinion, weisbuch2002meet} where attracting interactions only take place with like-minded individuals may be the most prominent. It has been demonstrated that these basic models and their generalizations, e.g.~\cite{hegselmann2015opinion,sirbu2017opinion,Redner2019}, allow to describe the emergence of opinion clusters or communities, including phenomena like radicalization, social bubbles and echo chambers~\cite{jager2005uniformity}. 

Additionally, biased assimilation models~\cite{Dandekar2013,Xia2020} have been developed based on the insight that linear weighted averaging of opinions cannot really lead to increasing polarization since people who hold strong opinions are likely to examine information in a biased manner. 
Further in~\cite{baumann2020modeling,baumann2021emergence, Leonard2022}, multi-dimensional opinion dynamics models have been proposed where individuals are drawn to the neutral opinion while at the same time reinforcing each other to more extreme opinions. Opinion interactions are saturated by a non-linear interaction function including tunable sensitivity to input. It has been demonstrated that these models allow for describing real-world political polarization~\cite{Leonard2021}.

The development of these opinion dynamics models has been taken up in the literature with a focus on understanding consensus and community building, or cluster formation. For simple linear models like the DeGroot model this can be studied analytically~\cite{degroot1974reaching}. For nonlinear opinion models like bounded confidence models this has been studied numerically~\cite{hegselmann2002opinion,weisbuch2002meet,carro2013role} or by considering the mean-field limit in terms of partial differential equations (PDEs), providing rigorous theorems on cluster numbers and size, effective timescales and bifurcations based on changing interaction parameters \cite{noisyHK,consensus2017,Delgadino_2021}. Moreover, it is understood how to extend the theory from agent-based models to mean-field limit models with stochastic partial differential equations (SPDEs)~\cite{helfmann2021interacting,conrad2022feedback}.

On the background of this progress in opinion dynamics in social networks, this article concentrates on opinion clusters as \emph{coherent structures}~\cite{froyland2010transport} meaning emerging groups of individuals whose opinions stay similar for rather long periods of time before eventually disintegrating or merging with other clusters. The notion of coherence implies temporal stability on timescales that are neither fast nor asymptotically long, indicating complex dynamical behavior with multiple characteristic timescales. We aim at understanding how, on these different timescales, traditional media and social media influencers may interact with individuals in order to adapt the overall opinion distribution according to their objectives, e.g., political agendas or economic interests. 

A few first steps towards opinion models under the impact of different types of agents have been taken: The influence of external sources on the opinion dynamics of individuals has been studied for example in~\cite{carletti2006make, mobilia2007role,hegselmann2015opinion}. These external sources are often regarded as media or charismatic leaders and have also been termed \emph{zealots} or \emph{stubborn agents} in the literature. More recently their influence has also been analysed in network-based opinion models~\cite{Baumann2020,Brooks_2020} but without explicitly taking coherence or multiple timescales into account. In~\cite{Ortiz2021}, the influence of a social group or clan at multiple timescales on the opinion formation on complex networks has been analyzed.
However, so far those agents are mostly considered to have constant opinions and relations to their followers that do not change in time.

\begin{figure}
    \centering
    \includegraphics[width = 0.7\linewidth]{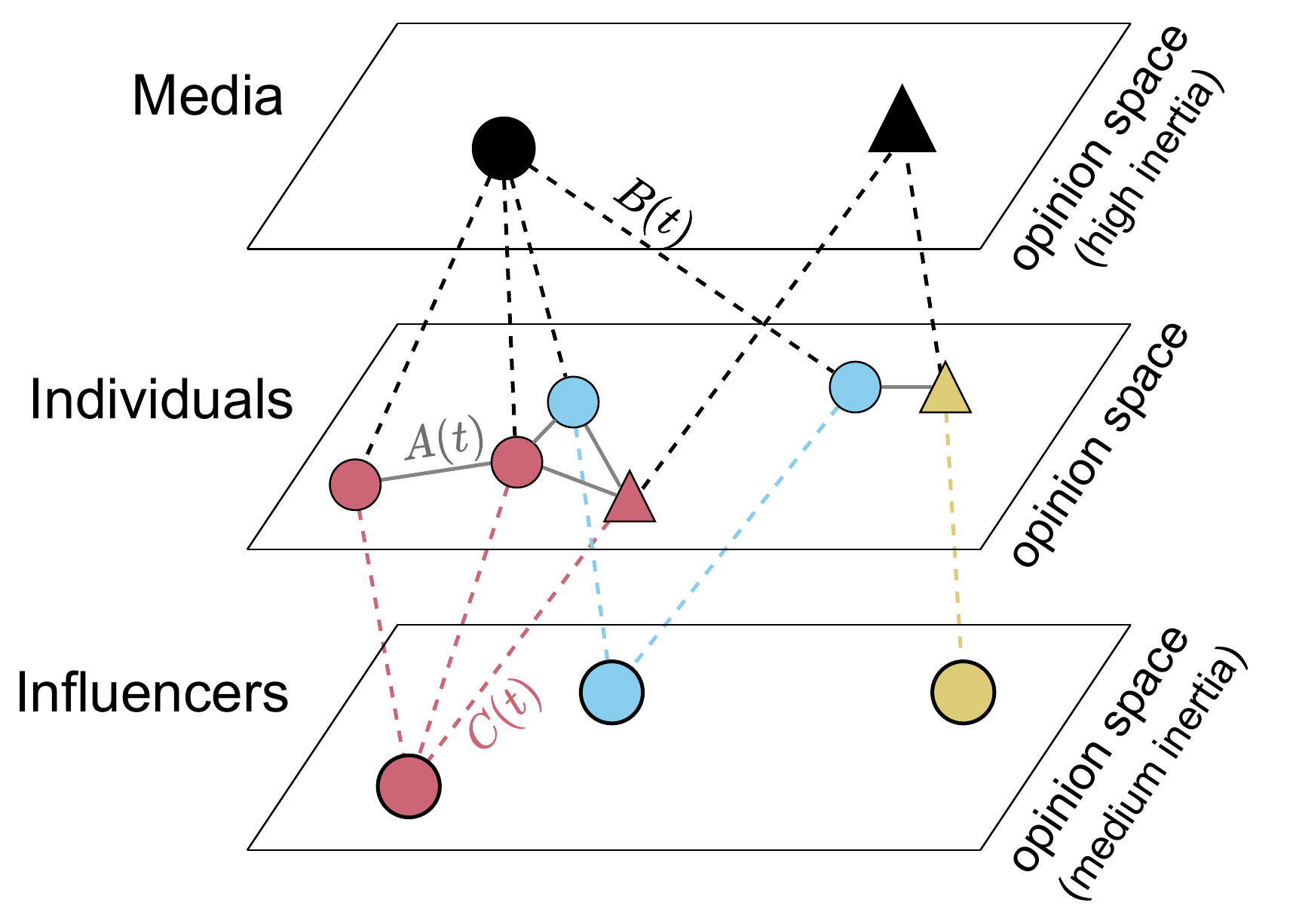}
    \caption{Structure of the agent-based model consisting of individuals, media and influencers. Each of the agents adapts its opinion in the continuous opinion space by interacting with its network neighbours. The network between individuals is given by $A(t)$ while the relations between individuals and media resp. influencers are defined by   $B(t)$ resp. $C(t)$. In the figure, the shape of an individual  indicates the medium they read 
and their color indicates the influencer they follow. Opinion changes happen on different time scales as determined by the inertia parameters where a higher inertia corresponds to slower opinion changes. }
    \label{fig:abm_structure}
\end{figure}

Here, we present an opinion dynamics model with a continuous opinion space that generalizes most of the available models (DeGroot, bounded confidence, biased assimilation,...) and complements individual opinion dynamics with traditional media and social media influencers that follow political agendas and economic interests, and that adapt their opinions on different timescales than the individuals, see Figure~\ref{fig:abm_structure}. 

The question of what role the media and influencers play on the public opinion distribution became increasingly complex. In the attention economy, the main goal of influencers and media is to receive more attention and increase their followership \cite{tafesse2021followers}, that is, to find strategies that achieve this goal in an \emph{optimal} way. This makes them themselves dynamic entities that can adjust their positions or agendas in order to achieve those goals \cite{gilardi2022social}. The effects these \emph{optimal strategies} in such an incentive structure could have on the overall opinion dynamics have not yet been fully understood.

Recent attempts to find control strategies of opinion dynamics consist of a variety of seemingly independent approaches. 
On the one hand, different heuristics have been developed to determine which agents should be targeted by an external controller to maximize their influence \cite{masuda2015opinion,Moreno2021,Li2021}.
On the other hand, in \cite{albi2015optimal,Kozitsin2022}, it is studied how \emph{external} controllers have to act to bring the opinion distribution of agents close to a specified target distribution, while,  in \cite{Li2021}, it is established how to control opinion distributions by (externally) interfering with the interaction network between individuals, and, in \cite{albi2017mean,gong2022crowd}, dynamic external controls for mean-field opinion models are considered.  

We will approach the design of optimal strategies in the following way: 
First, we show how to construct the mean-field model related to the proposed opinion dynamics model in the limit of large agent numbers and with finitely many influencers and media. This mean-field model takes the form of a system of partial differential equations for the individuals' opinion distribution  coupled to stochastic differential equations for the opinions of media and influencers. Then, we will demonstrate how this mean-field model can be used to study the effect of strategies in the attention economy. In particular, we will show how to derive optimal control schemes for counteracting agendas to influence opinion distributions. The advantage of the mean-field model compared to the ABM is that the computational cost is independent of the number of agents, and that the model is deterministic (when assuming that influencers and media are not affected by stochasticity), which, together, means that it is considerably easier to find insight into collective behavior or optimal strategies via the mean-field model.

\section*{Opinion model with influencers and media}

In the following we start by defining a general opinion model resulting from a large number of \emph{individuals} adapting their opinions through interaction with each other as well as due to the influence of a few specific agents with particular roles, namely traditional \emph{media} and social media \emph{influencers}. See Figure~\ref{fig:abm_structure} for the model components. 
We consider the situation of the early formation of opinions, which is of great importance in the accelerating public discourse \cite{lorenz2019accelerating}. Hence, we focus on the transient model dynamics to study the formation, splitting and merging of clusters, while the asymptotic regime is not of interest here. In this transient regime, individuals adapt their opinions on a fast time scale, while media and influencer agents change their opinion positions on a significantly slower time scale.

We consider individuals $i=1,\dots,N$ with opinions ${x_i(t)}$ lying in a continuous $d$-dimensional opinion space $D \subset \R^d$.\footnote{We regard the opinion space $D$ to be bounded with no flux boundary conditions.} The vector $\bm{x}(t) = (x_i(t))_{i=1}^{N}$ summarizes the opinions of all $N$ individuals.
In addition, we introduce $M$ highly influential agents that can be considered as media outlets with continuous opinions denoted by the vector $\bm{y}(t) = (y_m(t))_{m=1}^{M} \in D^M$, as well as $L$ highly influential influencers with opinions $\bm{z}(t) = (z_l(t))_{l=1}^{L} \in D^L$. 
To describe the real-world situation, we assume that there are much fewer influencers than individuals but still more influencers than media, i.e.,  $M<L \ll N$.   
When political opinions are modelled, the opinion space $D$ with $d=2$ could span the dimensions economic left $\leftrightarrow$ right and libertarian $\leftrightarrow$ authoritarian. When opinions with respect to acting against climate change are considered, two possible dimensions could be climate change believers $\leftrightarrow$ deniers and technocentric $\leftrightarrow$ ecocentric.

Relations among individuals such as friendship or connections on social media are defined through a binary adjacency matrix  $A(t)\in\{0,1\}^{N\times N}$ that can depend on time~$t$. The resulting network determines which individuals can interact (when there is an edge between two individuals) and which cannot (when there is no edge). When the edges are additionally weighted, such that $A(t)\in[0,1]^{N\times N}$, or directed, then the network can also describe the strength of a tie or the direction of social influence as in~\cite{degroot1974reaching}. 
Media and influencer agents on the other hand are assumed to only interact with those individuals that are their followers (readers/users). In the binary adjacency matrix $B(t)\in \{0,1\}^{N\times M}$ of medium-follower relations we store which individual follows which medium at time~$t$ while the binary adjacency matrix $C (t) \in \{0,1\}^{L\times N}$  defines the connections between individuals and influencers at time $t$.  

Matrices $A$, $B$ and $C$ represent the complex network structures of social interactions and can be derived for example from online social media or survey data. For the sake of simplicity, we will in our examples consider $A$ and $B$ to be constant on the chosen time scale and moreover an all-to-all connectivity between individuals, i.e. $A_{ij}=1$ for all $(i,j)$.  Additionally, we assume that each individual follows exactly one medium and one influencer, i.e., each row of $B$ and of $C$ contains exactly one entry. Since individuals usually change influencers very dynamically, we will propose an explicit model that defines how individuals are changing influencers in time and thus describes the evolution of $C$. This simplified setting allows us to better illustrate dynamical properties of the model and distinguish them from effects arising from the social ties. Our analytical results hold in the general case.

Individuals $i=1,\dots,N$ adapt their opinions in time according to the following stochastic differential equation (SDE)
\begin{equation}\label{eq:SDE_agents}
    dx_i(t) = F_i(\bm{x}, \bm{y},t)  dt + \sigma dW_i(t) 
\end{equation}
where $F_i$ defines the interaction force on individual $i$,  $\sigma>0$ gives the strength of noise and $W_i(t)$ are i.i.d. $d-$dimensional Wiener processes. The noise can, for example, model unknown external influences, uncertainties in the communication between individuals or free will. 
The interaction force on individual $i$ is given by a weighted sum of attractive forces  from all other connected individuals $j$ scaled by the parameter $a>0$ as well as attractive forces from the respective media scaled by the  parameter $b>0$ and from the respective influencer scaled by the parameter $c>0$
\begin{equation}\label{eq:F_i}
    \begin{split}
        F_i(\bm{x}, \bm{y}, \bm{z},t) =&  \frac{a}{\sum_{j'} w_{ij'}(t)} \sum_{j=1}^N   w_{ij}(t) (x_j(t)-x_i(t)) \\
         +& \frac{b}{\sum_{m} B_{im'}(t)}\, \sum_{m=1}^M B_{im}(t) \, (y_m(t)-x_i(t)) \\
         +& \frac{c}{\sum_{l'} C_{il'}(t)}\, \sum_{l=1}^L C_{il}(t) (z_l(t)-x_i(t)).
    \end{split}
\end{equation} 
The interaction weights between pairs of individuals $i,j$ are given by
\begin{equation}\label{eq:int_weights}
    w_{ij}(t) = A_{ij}(t)\,\phi\left(|x_j(t)-x_i(t)|\right),
\end{equation}
i.e., they depend on the network and on the opinion distance between the pair of individuals. The weights are normalized to ensure a unitary total contribution from all individuals. While the weight $w_{ij}$ is certainly zero when there is no edge between individuals $i$ and $j$, the weight depends on the opinion distance between $i$ and $j$ when there is an edge.\footnote{When an individual has an interaction weight $w_{ij}=0$ to all other other individuals, then the first term of the attraction force in~\eqref{eq:F_i} is assumed to be zero.} Possible choices for the non-negative, pair function $\phi(|x_j-x_i|)$ are given by:
\begin{itemize}
    \item $\phi(x) = \exp(-x)$ as in~\cite{mas2010individualization} that places exponentially more weight on close-by individuals,
    \item $\phi (x) = \1_{[0,d]} (x)$   as in bounded confidence models~\cite{weisbuch2002meet,hegselmann2002opinion,hegselmann2015opinion} that only allows interactions with other individuals that are within a radius $d$, here $\1_{[0,d]}$ is the indicator function on the set $[0,d]$,
    \item  $\phi(x) = 1$ as in the DeGroot model~\cite{degroot1974reaching} resulting in interactions irrespective of the opinion distance between individuals.
\end{itemize}  
The first two choices imply that individuals that are already close in opinion space excert higher social influence on each other (\emph{homophily}), while the third choice results in a weight irrespective of the opinions of the interacting individuals. 
\begin{remark}[Opinion differentiation and higher-order interactions]
Apart from the described opinion attraction (assimilation), it is also possible to model opinion repulsion (differentiation) between individuals. In~\cite{martins2010mass}, it is suggested to classify pairs of individuals~$(i,j)$ either as friends that attract in their opinions or as enemies that repel in their opinions. These pairs~$(i,j)$ will have a positive resp. negative edge weight $A_{ij}(t)$ and thereby can turn the first term of the force~\eqref{eq:F_i} from attraction to repulsion.  In~\cite{jager2005uniformity}, another approach that extends bounded confidence models is suggested:  pairs of individuals can not only become closer in their opinions when their opinion distance is smaller than the distance~$d$, but  they can also repel from each other when they are further apart than the distance $D>d$. By using the pair function
$\phi (x) = \1_{[0,d]} (x) - \1_{[D,\infty)}(x)$, this can be incorporated. But when including differentiation, the weights $w_{ij}(t)$ from~\eqref{eq:int_weights} can become negative and hence can no longer be normalized, instead one can for example normalize the force by the number of individuals, $N$, or by the number of neighbours of an individual.

Additionally, it is possible to describe higher-order interactions among more than two individuals at the same time with this model and thereby to more accurately describe group effects such as peer pressure, see e.g.~\cite{neuhauser2021opinion,conrad2022feedback}.
\end{remark}

Not only individuals change their opinions, but also media agents and influencers, however they adapt their opinions on a much slower time scale.
The resistance to rapid change is determined by the inertia parameter $\Gamma>1$ for media agents and by $\gamma>1$ for influencers. With $\gamma<\Gamma$ media agents are changing their opinions on an even slower time scale than influencers. In the limit when the parameters $\Gamma, \gamma$ diverge, the opinions of media and influencers become constant in time. There is a lot of research~\cite{carletti2006make, hegselmann2015opinion,bhat2020polarization,Brooks_2020} that studies interactions of individuals with constant agents (also termed \emph{stubborn agents} or \emph{zealots}) but to our knowledge adaptive media and influencers have not been studied so far.
In particular here media agents $m=1,\dots,M$ slowly adapt their opinions according to the SDE
\begin{equation}
    \Gamma dy_m(t) = f(\tilde{x}_m(t)-y_m(t)) dt + \tilde{\sigma} \tilde{dW}_m(t),
\end{equation} 
where the force function $f$ can be used to model nonlinear influence effects but is set to $f(x)=x$ subsequently, i.e.,  media agents are drawn in the direction of the average opinion of their followers
 $$\tilde{x}_m(t) = \frac{1}{\sum_{k} B_{km}(t)} \sum_{i=1}^N B_{im}(t) x_i(t).$$  
In the SDE,  $ \tilde{\sigma}>0$ gives the strength of noise on the opinion dynamics and $\tilde{W}_m(t)$ denote i.i.d. $d-$dimensional Wiener processes. 
Similar to media, influencers $l=1,\dots,L$ slowly change their opinions in the direction of their average followership according to the SDE
\begin{equation}
    \gamma dz_l(t) =g(\hat{x}_l(t)-z_l(t)) dt + \hat{\sigma} \hat{dW}_l(t),  
\end{equation} 
where the average opinion of followers is given by
$$\hat{x}_l(t) = \frac{1}{\sum_{k} C_{kl}(t)} \sum_{i=1}^N C_{il}(t)\,  x_i(t).$$
The noise strength is given by $\hat{\sigma}>0$, while $\hat{W}_l(t)$ denote i.i.d. $d-$dimensional Wiener processes. The function~$g$ can again be used to model nonlinear effects but is set to $g(x)=x$ in the following.

We have seen that influencers are similar to media agents but are usually more numerous and adapt their opinions on a faster time scale. To reflect that relationships on social media are more dynamic than to traditional media outlets, we further propose an explicit model of how  individuals can switch the influencer they are currently following. In particular, each individual~$i$ can at any time~$t$ switch its current influencer~$l'$  to another influencer~$l$  with a given rate $\Lambda_{m}^{\rightarrow l}(x,t)$ where $m$ is the medium of individual~$i$ and $x$ is the opinion of~$i$.\footnote{Note that the change rate does not depend on the current influencer of individual $i$ and that an individual can therefore also change to the  influencer it is currently following without any effect.} The rate could for example take the following form
\begin{equation}\label{eq:abm_rate}
    \Lambda_{m}^{\rightarrow l}(x,t) = \eta \, \psi(|z_l-x|) \, r\left(\frac{n_{m,l}(t)}{\sum_{m'=1}^M n_{m',l}(t)}\right)
\end{equation}
with scaling parameter $\eta>0$, pair function $\psi$, the link recommendation function~$r$ and~$n_{m,l}$ denoting the fraction of individuals that follow medium~$m$  and influencer~$l$. 
By setting the pair function for example to $\psi(x) = \exp(-x)$,  an individual has an exponentially higher rate to switch to an influencer that has a similar opinion than to an influencer with a very different opinion, i.e., there is homophily between influencers and individuals when connections are made. On social media platforms, link recommendation algorithms are often used to suggest new connections to users that  have the greatest potential to become established \cite{li2017survey,santos2021link}.
We incorporate link recommendation via the function~$r$ by assuming that individuals have a higher chance of switching to an influencer with a structurally similar followership. We measure the structural similarity of the followers of~$l$ to individual~$i$ by the ratio of followers that are connected to the same influencer (after switching) and medium as~$i$, this proportion is given by $\frac{n_{m,l}}{\sum_{m'} n_{m',l}}$.  We then assume that~$r$ is an increasing function on $[0,1]$, such as for example the ReLu function $r(x) = \max\{0.1, -1 + 2x\} $.

\begin{figure}
    \centering
    \includegraphics[width = 0.8\linewidth]{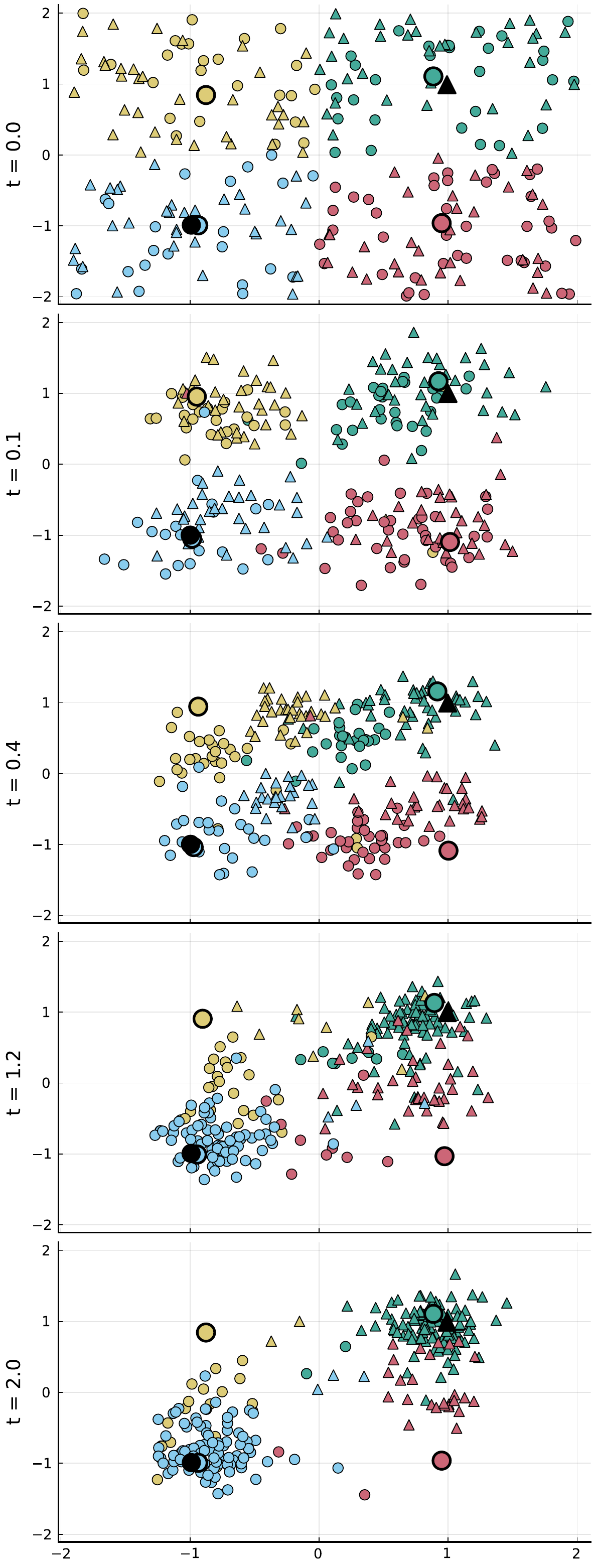}
    \caption{Realization of the ABM with media and influencers (Example 1). There are $2$ media agents marked in black, $4$ influencers indicated by large circles in $4$ colours, and 250 individuals that inherit the shape of the medium they read and the color of the influencer they currently follow.   Parameters: $a = 1$, $b=2$, $c=4$, $\sigma=0.5$, $\tilde{\sigma}=0$, $\hat{\sigma}=0$, $\Gamma=100$, $\gamma=10$, $A_{ij}=1$ for all $(i,j)$, $\phi(x) =\psi(x) = \exp(-x)$,   $\eta=15$.}
    \label{fig:abm_4inf}
\end{figure}

\begin{figure}[t]
    \centering
    \includegraphics[width = 0.88\linewidth]{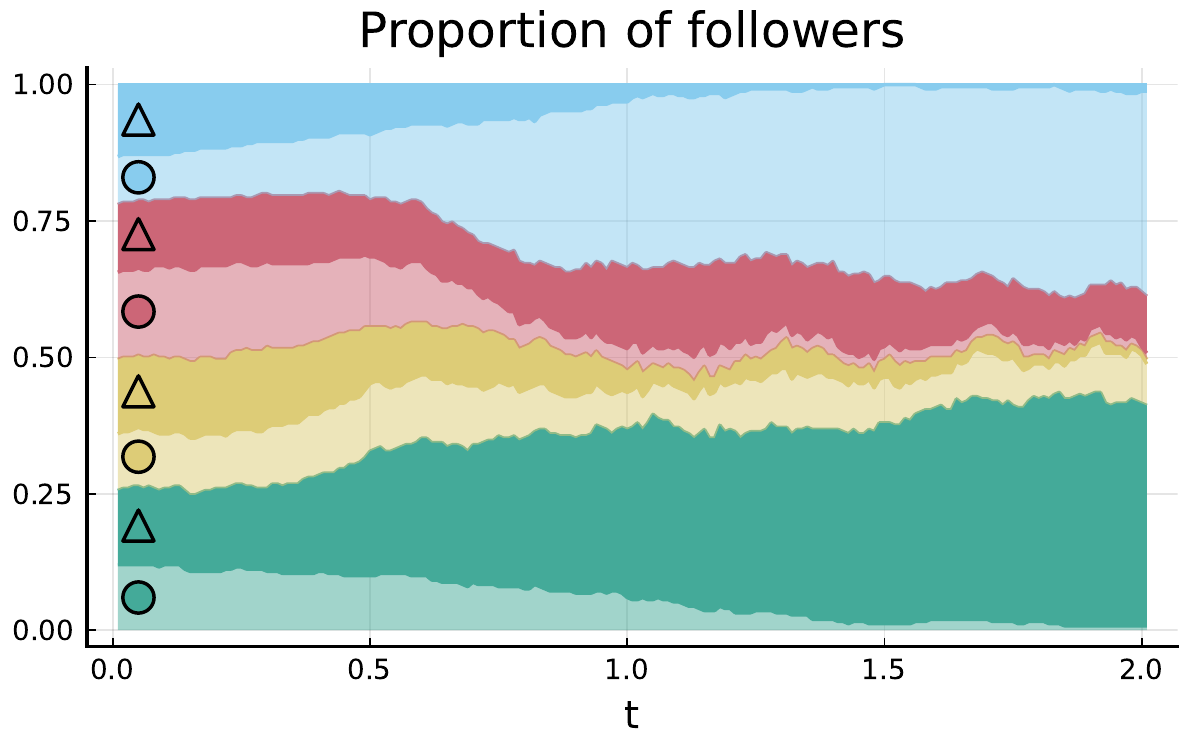}
    \caption{Stack plot showing how the proportion of individuals that follow a certain  influencer  and medium  (marked by the symbols on the left) evolve in time in Example~1.   }
    \label{fig:abm_followers}
\end{figure}
\subsection*{Example 1: ABM with media and influencers}
In Figure~\ref{fig:abm_4inf}, we show snapshots of one realization of the ABM with $250$ individuals,  $2$ media and $4$ influencers. The parameters are chosen to demonstrate how opinion clusters can form, split and merge. Initially at $t=0$, individuals are randomly distributed in opinion space and uniformly at random assigned to the $2$ media that are initially at $y_1(0)=(-1,-1)$ and $y_2(0)=(1,1)$. Individuals in each of the $4$ quadrants are assigned to a different influencer and the initial opinion of the influencer is set to the mean opinion of its followers. All individuals are interacting with each other, i.e., the network $A$ is fully-connected. 

When we let the model run, the strong attraction force to influencers ($c=4$) results in individuals quickly being attracted by their respective influencer and forming~$4$ clusters (compare with $t=0.1$). After some time, the $4$ clusters split further because individuals are also  attracted to their medium (compare $t=0.4$), s.t. individuals now form roughly~$8$ groups. Some individuals switch the influencer to a more suitable influencer, i.e., one that is closer in opinion space and whose majority of followers are connected to the same medium as the individual. They then get attracted to the new influencer ($t=1.2$), until finally ($t=2$) individuals have formed $2$ mixed clusters near the $2$ media opinions. 

In Figure~\ref{fig:abm_followers}, we show the evolution of the proportion of individuals that follow a certain influencer and medium. Around $t=0$, the proportions are roughly the same. But after $t=0.5$, individuals start switching their influencer (the medium cannot be switched), s.t. quickly the followers of each influencer are dominated by individuals following the same medium. Towards $t=2$, the proportion of individuals that follow the medium and influencer in the upper right corner (indicated as green triangles) and the  proportion of individuals that follow the medium and influencer in the lower left corner (shown by blue circles) dominate.

Also,  Fig.~\ref{fig:abm_4inf} shows that at $t=2$ most of the followers of the medium near $(-1,-1)$ (denoted by circles) follow the same influencer that is also near $(-1,-1)$ (colored in blue) while most followers of the medium near $(1,1)$ (denoted by triangles) also follow the influencer near $(1,1)$ (colored in green). The reason for this dominance is that for individuals it is favorable to be close to their medium and their influencer, otherwise they might switch the influencer to a more suitable candidate.
The two less suitable influencers in the upper left and lower right corner now  very slowly move towards their few remaining followers. On a larger time scale they will reach the clusters. Asymptotically, also the two clusters of individuals, media and influencers will approach another and merge. 
For different simulation runs, the agents behave qualitatively similar. But the individuals following the influencer initially at $(-1,1)$ and $(1,-1)$ are sometimes attracted to a different medium than in the shown simulation. 

The code for all examples is contained in the GitHub repository \url{www.github.com/LuzieH/SocialMediaModel}.

\subsection*{Rich dynamical behaviour} 
The situation studied in Example 1 is somewhat symmetric and idealistic. For other initial configurations and other choices of parameters and pair functions~$\phi$ different forms of complex dynamical behavior emerge. This ranges from stable opinion clusters centered around "their" influencers (for $\phi (x) = \psi(x) = \1_{[0,d]} (x)$ and no media) to a complex interplay of cluster formation and reformation for several influencers with smaller and larger inertia~$\gamma$, and a less symmetric configuration as in Example 1.  The dynamics can become even richer, if the interaction network $A$ is already exhibiting clusters. Moreover, different initial conditions lead to different transient dynamics.

\section*{Partial mean-field (opinion) model}\label{sec:pde}
\begin{figure}
    \centering
    \includegraphics[width = 1\linewidth]{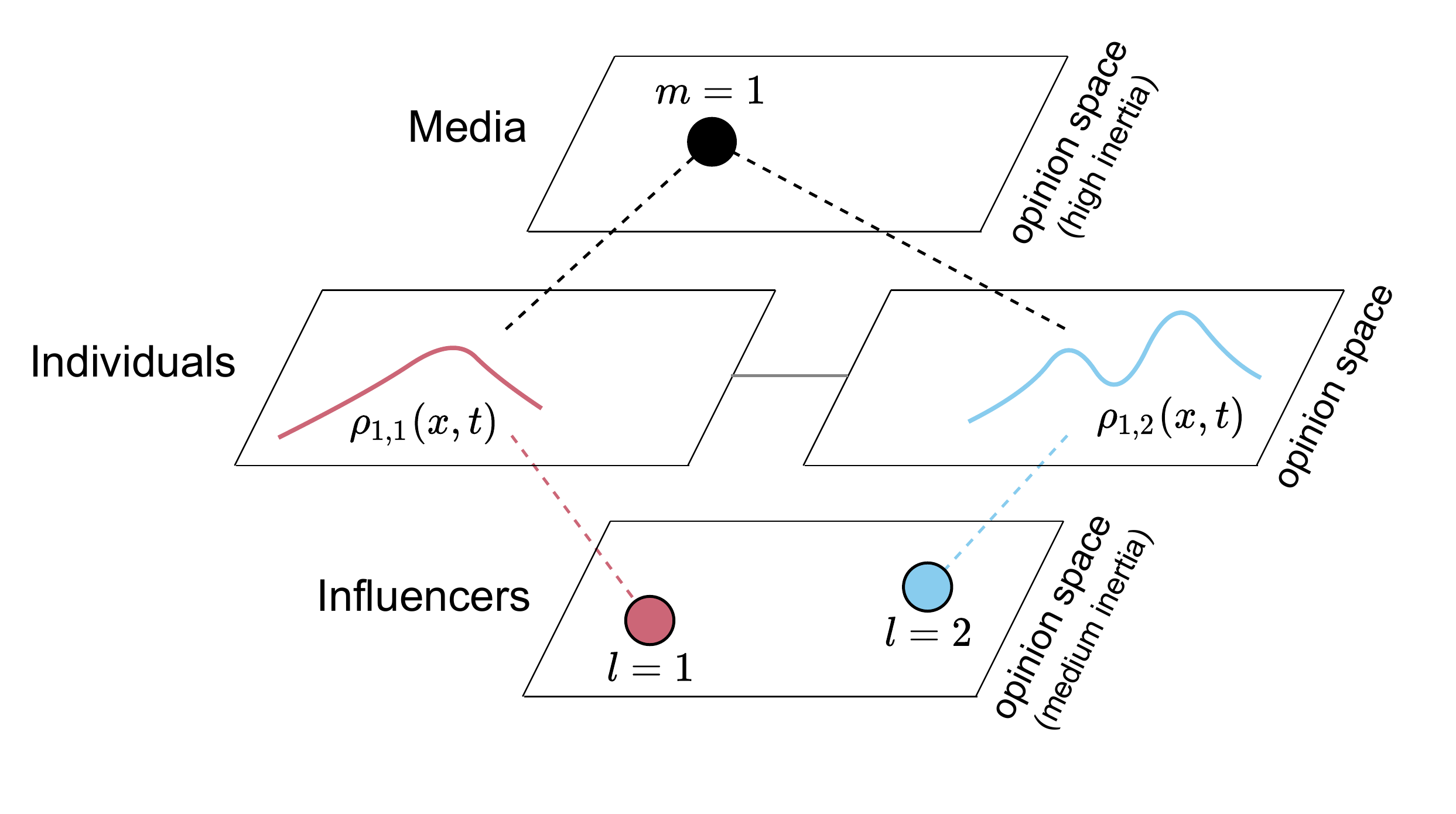}
    \caption{Structure of the partial mean-field model of distinct  media agents (indicated by black circles)  and influencers (shown as colored circles)  in interaction with the opinion distributions of individuals. The opinion distribution of individuals that follow a certain medium $m$ and influencer $l$ is given by~$\rho_{m,l}(x,t)$. 
}
    \label{fig:pde_structure}
\end{figure}

For situations with many individuals but few influencers and media, one can derive the mean-field limit  by a partial differential equation (PDE) that describes the opinion dynamics of individuals in the limit of infinitely many individuals~\cite{noisyHK,consensus2017,goddard2022noisy} but is usually already a good approximation to the dynamics for finitely many individuals. Since here the number of influencers and media is still small and finite, their dynamics are still best described by SDEs but now coupled to PDEs for the evolution of the opinion distributions of individuals. We therefore also call the model a \emph{partial} mean-field  model, compare Figure~\ref{fig:pde_structure} for the structure of the model. The coupled system of PDEs and SDEs is not only computationally advantageous~\cite{helfmann2021interacting} since the computational effort no longer scales with $N^2$ (due to the expensive computations of pair-wise distances of individuals in the ABM) but with the number of spatial grid cells and independently of $N$. Additionally, the model is  conceptually easier to study for example to find critical parameters~\cite{noisyHK,consensus2017,Delgadino_2021} or to use the partial mean-field model to control the evolution of the opinion distribution through different influencer and media strategies (see next section).

For the sake of simplicity, we subsequently assume a  fully-connected interaction network between individuals, such that $A_{ij}=1$ for all pairs of individuals $(i,j)$\footnote{Without this assumption, one would need to derive mean-field dynamics for interacting agents on  realistic complex networks that have a network limit in terms of graphons~\cite{HayatoChiba} or graphops~\cite{kuehn2020network,gkogkas2022graphop}, but here we will make the standard assumption of a fully-connected network between individuals.}, and that each individual follows exactly one medium and one influencer at a time and only influencers can be switched at the rate given in~\eqref{eq:abm_rate}.

Then let us define the \emph{empirical distribution} of individuals that follow  medium $m$ and influencer $l$ at time $t$ by the sum of Dirac Delta distributions $\delta$ placed at the individuals' opinions
\begin{equation}
    \rho_{m,l}^{(N)} (x,t) = \frac{1}{N} \sum_{\substack{i: B_{im}=1,\\ C_{il}(t)=1}}\delta(x-x_i(t)). 
\end{equation}  
This distribution describes the stochastic opinion instances at a given time~$t$ and   integrates to
$$\int_D \rho_{m,l}^{(N)} (x,t) dx =: n^{(N)}_{m,l}(t),$$
the proportion of individuals that follow medium $m$ and influencer $l$. 

It can be shown (see Supplementary Material) that  as $N\rightarrow \infty$, the empirical distribution  $\rho_{m,l}^{(N)} (x,t) $ can be replaced by the limiting distribution $\rho_{m,l}(x,t)$ solving the following PDE on the domain $D$ 
\begin{equation}\label{eq:agent_pde}
\begin{split}
    \partial_t\rho_{m,l}&(x,t)  =  \frac{1}{2} \sigma^2 \Delta \rho_{m,l}(x,t) - \nabla\cdot\left(\rho_{m,l}(x,t) \, \mathcal F(x,y_m, z_l, \rho) \right)\\
&  + \sum_{l'\neq l} \left(
     - \Lambda_{m}^{\rightarrow l'}(x,t)\,  \rho_{m,l}(x,t) + \Lambda_{m}^{\rightarrow l}(x,t)\,  \rho_{m,l'}(x,t)
    \right)
\end{split}
\end{equation}
for each $m=1,\dots,M$ and $l=1,\dots,L$, where $\nabla\cdot $ denotes the divergence operator and $\Delta$ the Laplace operator on opinion space. The PDE is accompanied by boundary conditions ensuring that the number of individuals in the system is conserved. 

The changes of $\rho_{m,l}$ are governed by three processes:
(i)~The first term on the RHS of the PDE is responsible for the stochastic diffusion of opinions. (ii)~The divergence term models the interaction of the distribution $\rho_{m,l}$ with the overall distribution of all individuals $\rho=\sum_{m,l} \rho_{m,l}$ as well as with the respective medium $y_m$ and the influencer $z_l$ according to the attraction force at opinion $x$
\begin{equation}
    \begin{split}
        \mathcal F(x,y_m, z_l, \rho) =& a \, \frac{\int_D \rho(x',t) \phi(|x' - x|) (x'-x)\, dx'}{\int_D \rho(x',t) \phi(|x' - x|)\, dx'} \\
        & + b\, (y_m(t)-x) + c\, (z_l(t)-x).
    \end{split}
\end{equation} 
 (iii)~The last term of the PDE is responsible for the mass exchange  between different distributions due to individuals switching the influencer away from~$l'$ or towards~$l$. Note that due to the second and third term, the PDE is non-local.

The PDEs are coupled to the SDEs of the media $m=1,\dots,M$
\begin{equation}\label{eq:limit_media}
    \Gamma dy_m(t) =  (\tilde{x}_m(t)- y_m(t)) dt + \tilde{\sigma} \tilde{dW}_m(t)
\end{equation}  
where the average opinion of followers is now given by $$\tilde{x}_m(t) = \frac{ \sum_{l=1}^L\int_D    x \rho_{m,l}(x,t) dx}{\sum_{l=1}^L n_{m,l}(t)}$$
with $ n_{m,l}(t)$ denoting the limit of $ n_{m,l}^{(N)}(t)$,
as well as to the SDE dynamics of influencers $l=1,\dots,L$
\begin{equation}\label{eq:limit_inf}
    \gamma  dz_l(t) = ( \hat{x}_{l}(t)- z_l(t) )dt + \hat{\sigma} \hat{dW}_l(t),
\end{equation}
with $$\hat{x}_{l}(t)= \frac{ \sum_{m=1}^M\int_D    x \rho_{m,l}(x,t) dx}{\sum_{m=1}^M n_{m,l}(t)} $$  denoting the average opinion of individuals that follow influencer $l$.

\begin{figure}[t]
    \centering
    \includegraphics[width = 1\linewidth]{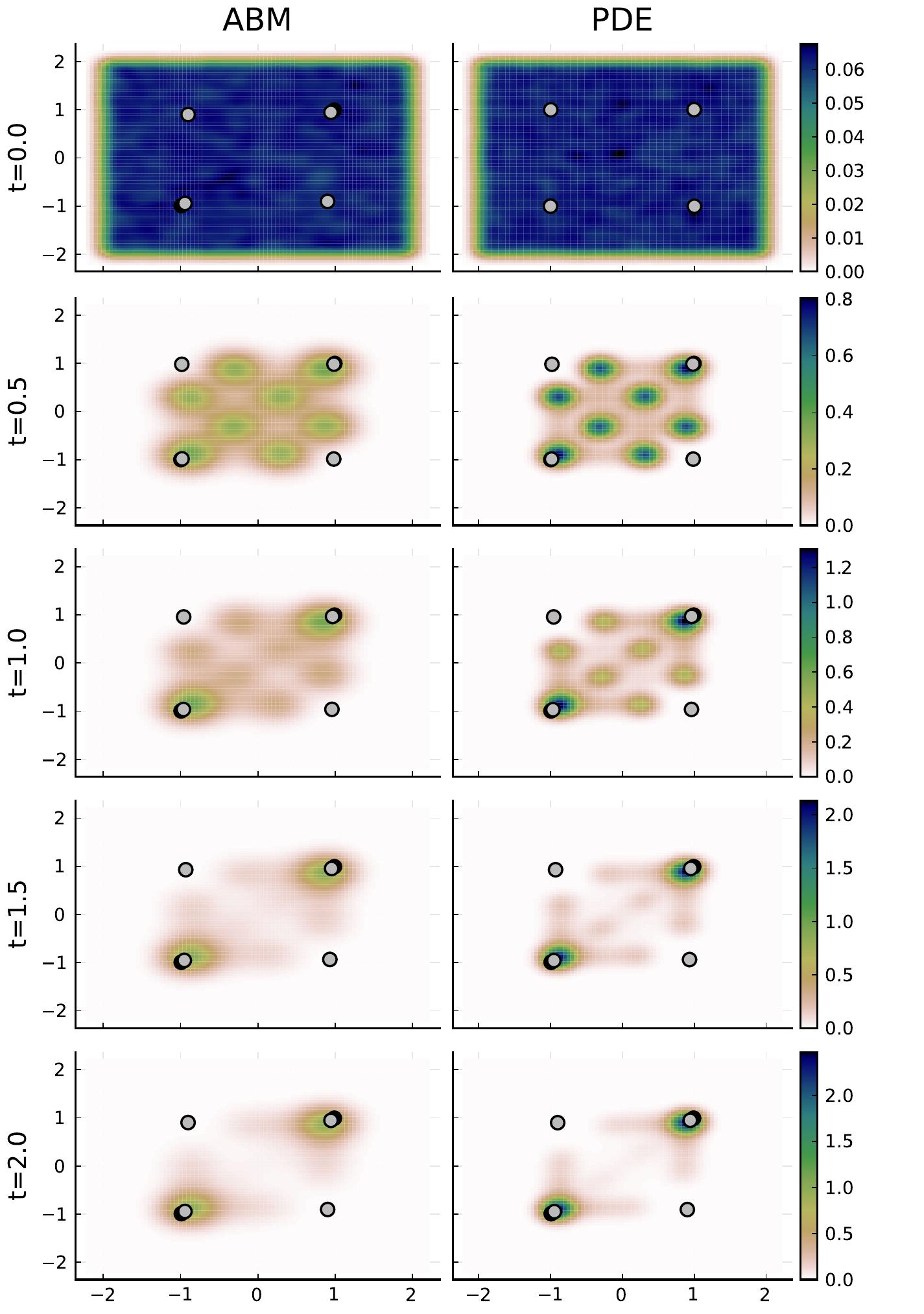}
    \caption{Mean configuration of the ABM (left column) and partial mean-field PDE (right column) over 1000 simulations with influencers and media. Influencers are now marked by grey circles, and media agents by black circles. The opinion distribution of all individuals (i.e., independent of the media/influencer they follow) is shown with a heat plot. Parameters as in Example 1.}
    \label{fig:ensemble_mean}
\end{figure}

\subsection*{Example 2: Comparison on ABM and partial mean-field dynamics}
In Fig.~\ref{fig:ensemble_mean}, we show a comparison of ABM and partial mean-field simulations averaged over 1000 realizations, the parameters are as in Example~1.
To compare the ABM configurations against the opinion distributions in the partial mean-field, the opinion distributions resulting from the ABM are visualized via Kernel density estimation with Gaussian kernels. Even though the PDE is deterministic, we initialized it with random initial distributions given by the Kernel density estimation of $250$ individuals placed uniformly at random in the domain. Different realizations of these random initial conditions lead to different transient dynamics and we therefore also averaged the partial mean-field simulations. The comparison shows that the results of the ABM and the partial mean-field are very consistent already for $N=250$ individuals in the system.

\section*{Strategies in the partial mean-field model}\label{sec:inf_strat}
In the previous section we have stated the partial mean-field model as a reduced model describing the opinion dynamics of infinitely many individuals coupled to the opinion dynamics of finitely many media and influencers. Assuming deterministic opinion changes of influencers and media  (i.e., $\hat{\sigma}=\tilde{\sigma}=0$), and a deterministic initial opinion distribution, the partial mean-field model is deterministic and faster to solve than the ABM, allowing us to  study the effect of different (optimal) influencer and media strategies. 

\begin{figure}[htb!]
    \centering
    \includegraphics[width = 0.8\linewidth]{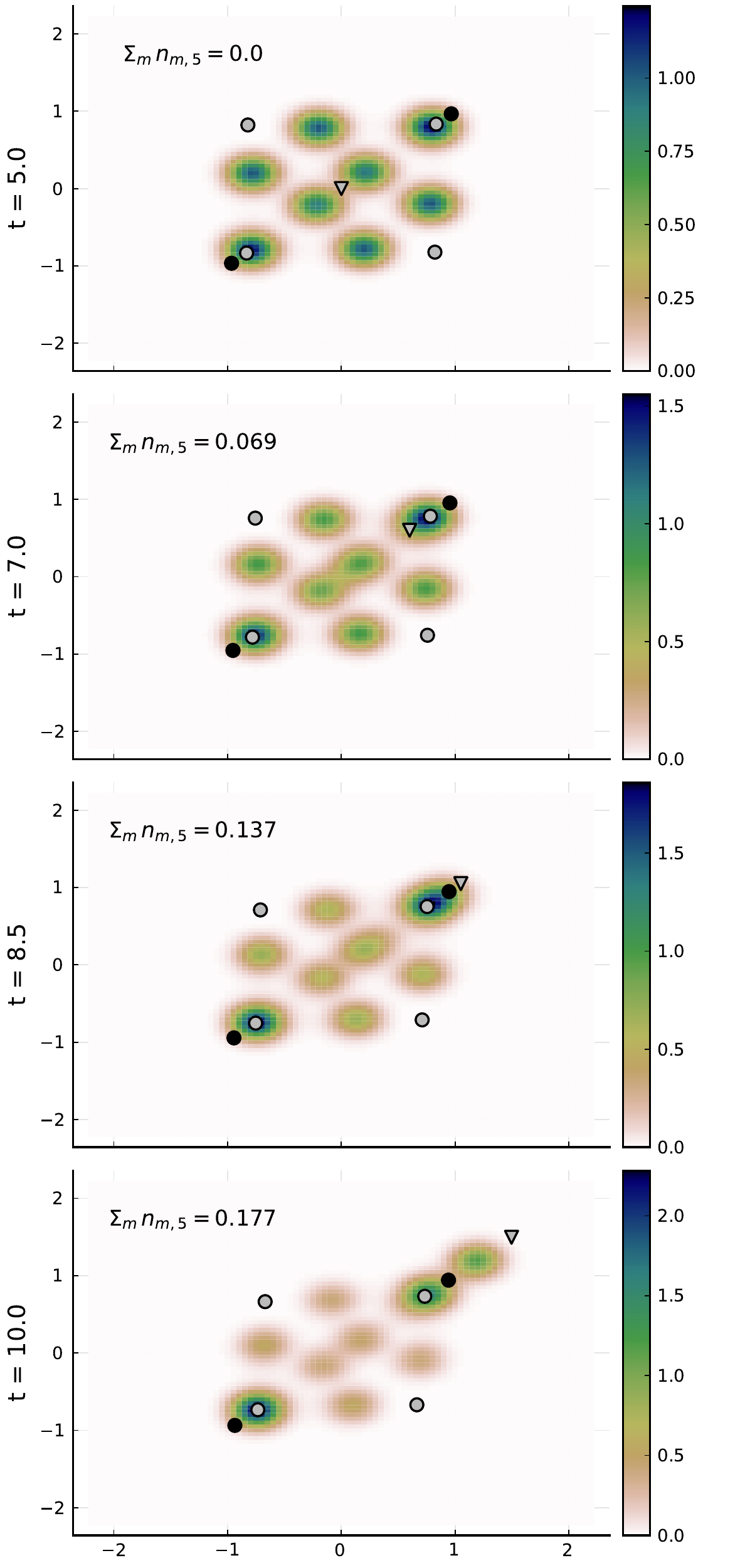}
    \caption{Opinion distribution $\rho$ evolving in time (Example 3) started from a deterministic uniform opinion distribution at $t=0$. At time $t=5$, a 5th influencer (marked by a triangle) is inserted into the system and moves at a constant pace in the direction of the right upper corner while collecting new followers. Parameters as in Example~1, except $\eta=1$.}
    \label{fig:steering}
\end{figure}

\subsection*{Increasing followership}
In the version of our opinion dynamics model as specified above, influencers adapt their opinions in the direction of the average opinion of their followers. This is a very simple strategy for influencers trying to keep their followers attached to them \cite{tafesse2021followers}. 
Other strategies to keep and even increase the number of followers might be more fruitful. In the following example we discuss a simple strategy for a new influencer with initially zero followers to substantially increase its followership. 

\subsection*{Example 3: Strategy of an influencer to increase followership}
We consider the dynamics as before in Example 1 except with $\eta=1$.
With this choice of parameters, influencers strongly affect and attract individuals (since $c>a,b$) and  individuals only slowly change influencers (due to $\eta$ being small). In this way,  once individuals follow  a certain influencer, they will remain with that influencer for some time. 

At time $t=5$  we insert an additional influencer into the system with opinion~$(0,0)$ and   zero followers. This new influencer then moves at a constant speed to the opinion $(1.5,1.5)$ during the time interval $[5, 10]$.
In Fig.~\ref{fig:steering}, we show snapshots of the realization. The inserted influencer $l=5$ quickly collects and attracts many followers behind. Starting with initially zero followers, the final followershare is $$\sum_{m=1}^2 n_{m,5}(10.0)\approx 0.18.$$ Moreover the time-averaged proportion of followers of influencer $l=5$ over the interval $[5,10]$ comes out to be approximately $$\frac{1}{5}\int_{t=5}^{t=10} \sum_{m=1}^2 n_{m,5}(t)\, dt \approx 0.10.$$ 
Thus the added influencer could substantially increase its followership by moving to an extreme opinion position. Even a new cluster of followers has formed near the influencer.   

\subsection*{Optimal counteraction}
More generally, the partial mean-field model also allows to apply optimal control techniques in order to derive the best strategy for a single influencer or a medium to achieve a certain objective, such as maximizing the number of their followers or optimally counteracting the goal of another agent.

Let $\varrho(u)=(\rho_{m,l}(t))_{m=1,\ldots,M, l=1,\ldots,L}$, the matrix of all different opinion distributions on the time interval~$[0,T]$, satisfy Eqs.~(\ref{eq:agent_pde}), (\ref{eq:limit_media}), and (\ref{eq:limit_inf}), except for the chosen controlled influencer or medium. When we are interested in the strategy for an influencer agent $l=l^\star$, we let the control $u$ determine the behaviour of the $l^\star$th influencer, i.e., $z_{l^\star}=u$, when on the other hand we are searching for a media strategy, we control the $m^\star$th medium with $y_{m^\star}=u.$ 
We can then  express the aim of influencer $l^\star$ to maximize its temporally aggregated followership with the objective function
\[
    \mathcal{J}^\varrho = \int_{0}^T   \sum_{m=1}^M n_{m,l^\star}(t)  \,dt,
\]
while maximizing solely the final followership means dropping the integral and taking the integrand at time $t=T$.
In contrast, if the goal of influencer $l^\star$ or medium $m^\star$ is to counteract the maximization of followership of another influencer $l'$, this can be achieved by maximizing the objective 
\[
    \mathcal{J}^\varrho = -  \int_{0}^T   \sum_{m=1}^M n_{m,l'}(t)  \,dt.
\] 
When the control determines the behavior of influencer $l^\star$ or medium $m^\star$, the control $u$ needs to satisfy certain restrictions. That is, the control function~$u$ has to come from a set of admissible controls~$U$. On the one hand, $U$ is restricted by the necessary domain constraints $u(t)\in D$ for almost all $t\in[0,T]$. On the other hand, with every control~$u$ certain costs are associated. Control discontinuities or even chattering controls could, e.g., increase the risk of the control activities being detected, and countermeasures taken. 
Such concerns can be included as a penalty term 
in the objective, e.g., by 
\[
    \mathcal{J}^p
    = \alpha |\partial_t u|^2_{L^2(\mathopen]0,T\mathclose[)},  
\] 
where  the parameter $\alpha$ regulates the penalty for large opinion changes. Then, the optimal control problem has the general form 
\begin{equation}\label{eq:control-problem}
\max_{u\in U}\;\mathcal{J}^\varrho - \mathcal{J}^p .
\end{equation}
After an appropriate control discretization with, e.g., piecewise polynomials, the resulting nonlinear programming problem can be solved by derivative-free methods like Nelder-Mead or by more efficient inexact gradient descent~\cite{NocedalWright1999} or stochastic approximation methods. Sufficiently accurate gradient evaluations can be obtained by finite differencing the PDE/SDE forward equations for $\varrho(u)$ or by Feynman-Kac type gradient sampling.

\subsection*{Example 4: Influencer counteraction}
\begin{figure}[htt!]
    \centering
    \includegraphics[width = 0.8\linewidth]{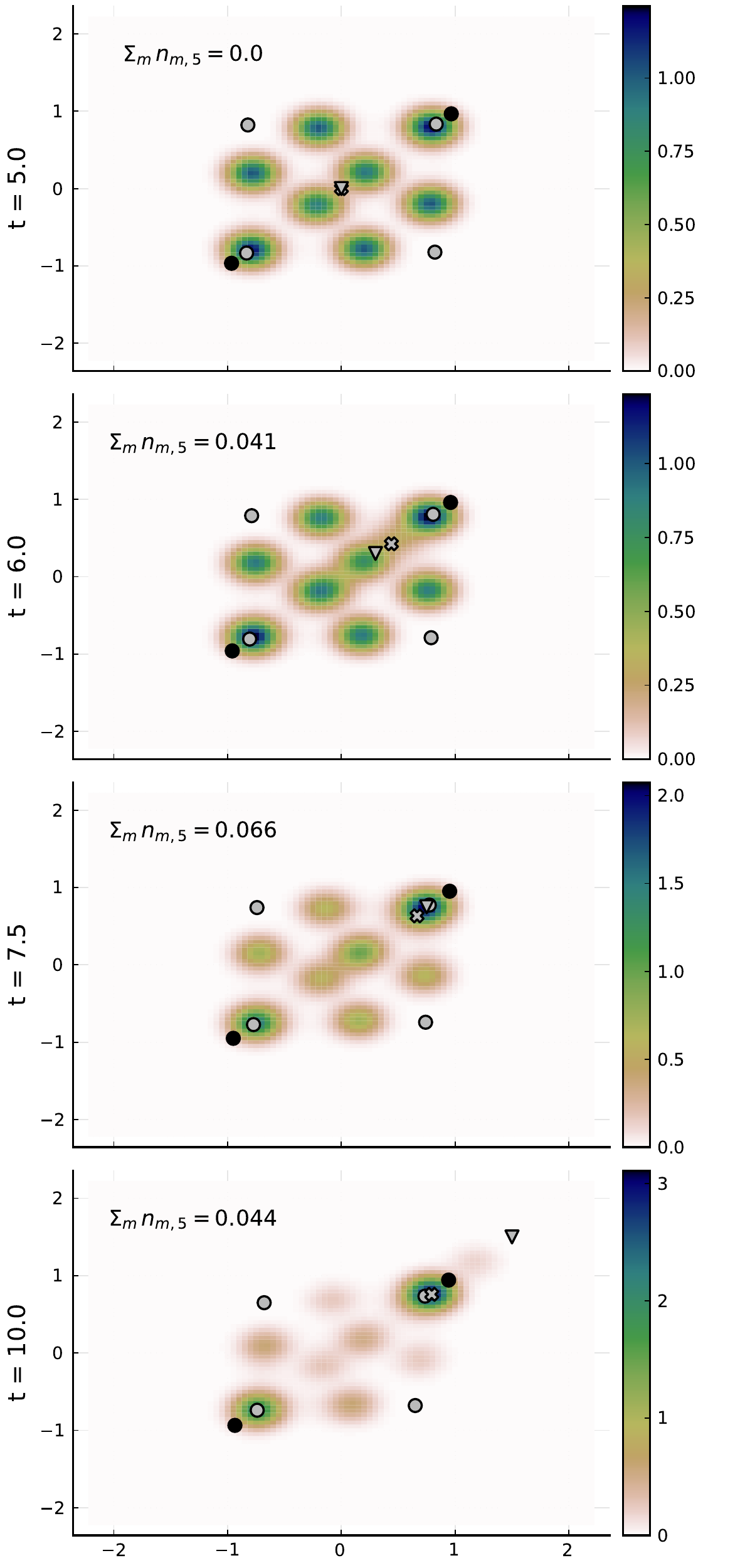}
    \caption{Opinion distribution $\rho$ evolving in time (Example 4) started from a uniform distribution at $t=0$. At time $t=5$, a 5th influencer (marked by a grey triangle) is inserted into the system and moves at a constant pace in the direction of the right upper corner.  Simultaneously, a $6$th influencer  (marked by a cross) starts and moves in a optimal way to prevent influencer $l'=5$ from collecting many followers. Parameters as in Example~4.}
    \label{fig:counterinf}
\end{figure} 
We are interested in the optimal counteraction of influencer $l^\star=6$   when influencer $l'=5$ moves at a constant speed to the upper right corner~$(1.5,1.5)$ (as in Example~3) and thereby drastically increases its followership. We therefore want influencer $l^\star$ to move in an optimal way to satisfy 
\begin{equation}\label{eq:obj_counter} 
\max_{u\in U}\; \left(- \int_{5}^{10}   \sum_{m=1}^2 n_{m,5}(t)  \,dt - \alpha |\partial_t u|^2_{L^2(\mathopen]5,10\mathclose[)}\right)
\end{equation}
with $\alpha=0.05$. That is, the goal is to minimize the followershare of influencer $l'$, while avoiding conspicuously drastic opinion changes. The control set $U$ contains all functions with $u(t)\in D$ and with piecewise constant velocities that only change at $3$ chosen time points.

The optimal strategy is shown in Fig.~\ref{fig:counterinf} and results in an average followershare of influencer $l'$ of  $$\frac{1}{5}\int_{t=5}^{t=10} \sum_{m=1}^2 n_{m,5}(t)\, dt \approx 0.05,$$  which is significantly smaller compared to the value $0.1$ that was obtained in Example 3 without counteraction. Note also, that the final followershare at time $t=10.0$ was decreased from $0.18$ without counteraction to $0.04$ with counteraction. 
The snapshots in Fig.~\ref{fig:counterinf} show that the optimal counterstrategy of influencer $l^\star=6$ is to steel followers of influencer $l'=5$ by moving along; it thereby stabilizes the opinion cluster of individuals. Note that according to the objective specified in~\eqref{eq:obj_counter}, this  counteraction comes out to be more effective than the movement in the opposite direction of influencer $l'=5$. 

\subsection*{Example 5: Media counteraction}
\begin{figure}[htb!]
    \centering
    \includegraphics[width = 0.8\linewidth]{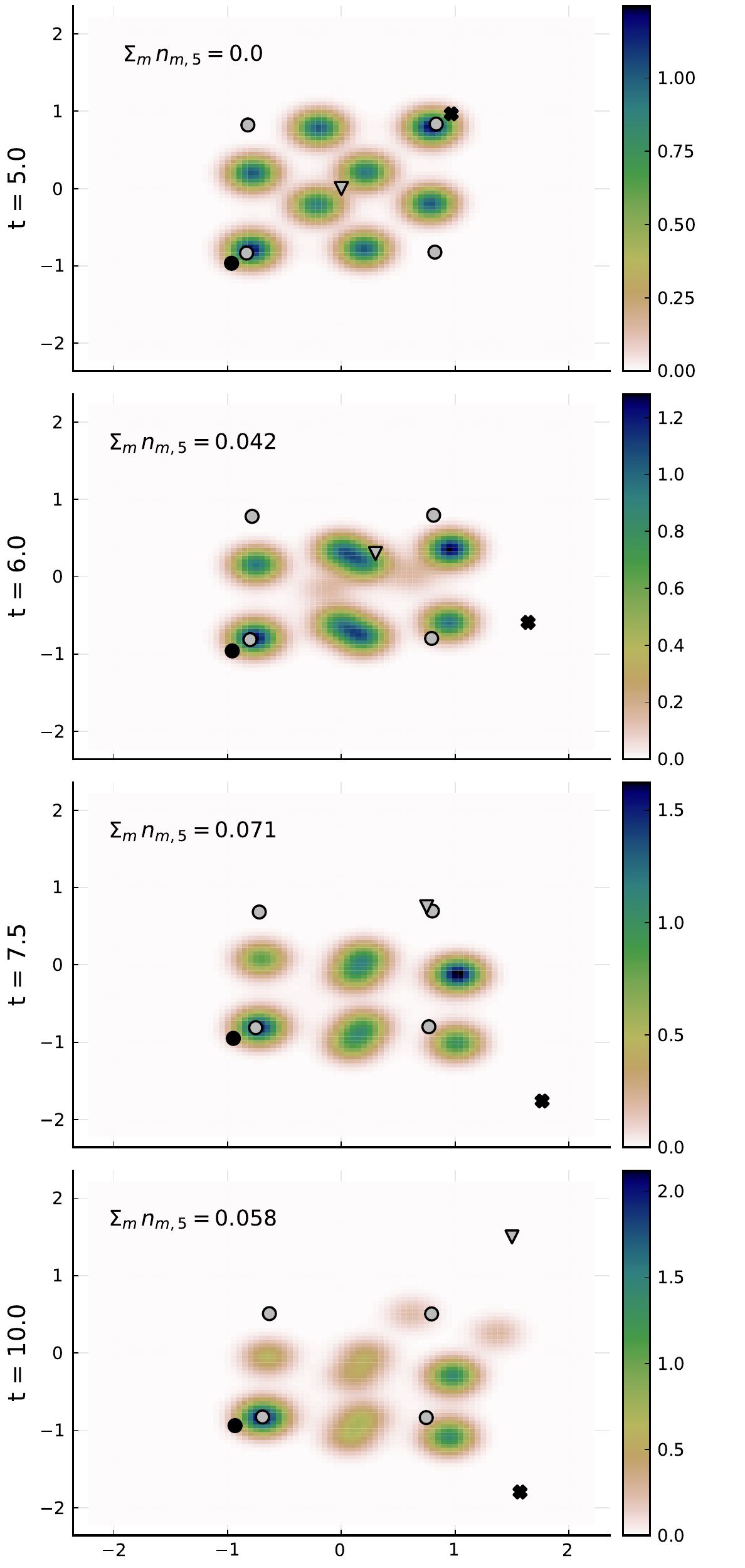}
    \caption{Opinion distribution $\rho$ evolving in time (Example 5) started from a uniform distribution at $t=0$. At time $t=5$, a 5th influencer (marked by a triangle) is inserted into the system and moves at a constant pace in the direction of the right upper corner, simultaneously, the media in the upper right corner (marked by a cross) moves in a optimal way to prevent influencer $l'=5$ from collecting many followers. Parameters as in Example~4.}
    \label{fig:countermed}
\end{figure} 
Next, we study the optimal counterstrategy of the 
media agent $m^\star=2$ in the upper right corner when trying to minimize the followershare of influencer $l'=5$, analogously to the previous example with the objective function given by~\eqref{eq:obj_counter}. 
The resulting optimal strategy is shown in Fig.~\ref{fig:countermed}. 
By changing the opinion drastically from the upper right to the lower right corner, the media agent manages to decrease the average followershare of influencer $l'=5$  down to $$\frac{1}{5}\int_{t=5}^{t=10} \sum_m n_{m,5}(t)\, dt \approx 0.06.$$  Note that since the system dynamics are symmetric wrt. the diagonal axis $x=y$, an equivalently good strategy would be to move to the upper left corner. The counterstrategy of the media agent is very different to the counteraction of an influencer (given in Example 4). This is because influencers compete for followers and can steal influencers from each other while media agents can only make an opinion topic unattractive to individuals by changing the topic.

\section*{Conclusion}

In this work we provided new mathematical means for the systematic study of how traditional media and influencers might impact coherent structures of the public opinion distribution as modelled by opinion dynamics models. It does neither claim to describe opinion shifts in real-world social networks nor does it state anything about influencing the opinion of an individual human being. It is still an idealized model and, like most contributions to the field, can neither claim to describe human opinion formation processes realistically, nor to be validated with observational data and controlled sociological experiments. However, it may help to describe how shifting individual perspectives and social exchange lead to archetypal states of public opinion distribution like coherent opinion clusters. By providing a strategy for understanding how influencing the public opinion distribution may be done and counteracted in an optimal way, this work may help us to understand how to face current challenges of opinion polarization in complex scenarios. 

In particular, our model demonstrates the impact influencers and media might have on the opinion distribution by creating stable and coherent opinion clusters that are drawn to the positions of the respective influencer or media. We also see that when influencers move to extreme positions, they fragment the opinion landscape and gain more followers and more attention in the competitive scenario of social media. Hence, these results suggest that those competitive goals could contribute to a polarized and fragmented debate.
However, our work also offers avenues towards solutions through optimal counteracting those attempts via other influencers (or media). These counteracting influencers can stabilize the existing opinion landscape against extreme influencers by strengthening existing opinion clusters. 
Empirical work needs to be done to validate those strategies and interplays that our model is suggesting, potentially through large-scale field experiments on social media. Our theoretical work potentially delivers a good basis for developing such alternative strategies for influencers in the current system to stabilize online discourse.

In this work we did not explore the full generality of the opinion dynamics model for different interaction terms, parameters and different influencer and media strategies. 
Instead we focused in this work on the feasibility by designing the pipeline from the inclusion of influencers and media via the demonstration of the emergence of temporarily coherent opinion clusters to the options for influencing the public opinion by optimally chosen strategies. The results provided do not include any analysis of the dynamical patterns such as numbers, birth and decay of opinion clusters, that the model may exhibit for different parameter combinations. This was beyond the focus and scope of this study and will be left to future investigations.

\paragraph{Acknowledgments} We would like to thank Jobst Heitzig for his insights into the opinion model, Ana Djurdjevac, Stefanie Winkelmann and Alberto Montefusco for their comments on mean-field models, Alexander Sikorski for discussions on PDE discretizations and insights into code speed-ups, and Martin Weiser for valuable conversations  on optimal control theory. This work has been partially funded by the Deutsche Forschungsgemeinschaft (DFG) under Germany’s Excellence Strategy through grant EXC-2046  \emph{The Berlin Mathematics Research Center MATH+} (project no.  390685689).  P.L.-S. acknowledges financial support from the Volkswagen Foundation (grant ‘Reclaiming individual autonomy and democratic discourse online: How to rebalance human and algorithmic decision-making’).  

\onecolumn
\section*{Supplementary Material}
\appendix
In the following, we will outline the derivation of the partial mean-field model for the limiting dynamics of infinitely many individuals, while keeping the number of media and influencers fixed. 
\subsection*{PDE Derivation}
We start by deriving the mean-field PDEs for the opinion dynamics in the limit of infinitely many individuals.
Related mean-field PDEs have already been derived in the context of other opinion models~\cite{wang2017noisy,consensus2017,goddard2022noisy}.
\paragraph{Assuming constant influencer-follower connections.}

For the derivation, we will assume that the network between individuals is fully-connected, i.e. that $A_{ij}=1$ for all $i,j$, and that each individual  follows exactly one medium and one influencer. Further in this first part,  we will suppose that individuals do not change the influencer they are following, i.e. the matrix of influencer-follower relations remains constant $C(t) \equiv C$. In a next step we will relax this assumption. 
For the purpose of this derivation, we add ${(N)}$ to the model quantities when the number of individuals is still finite. 

Since there are $M$ media and $L$ influencers in the system, we have $M\times L $ different types of individuals depending on which combination of a medium and an influencer a particular individual is following. Since the network between individuals is fully-connected, individuals that follow the same influencer and medium can be considered as identical (exchangeable). For that reason, we can switch from  a description in terms of labeled individuals to a description in terms of empirical distributions. 
Let us define the empirical distribution of individuals that follow medium $m$ and influencer $l$ by 
\begin{equation}\label{eq:emp}
\rho_{m,l}^{(N)} (x,t) := \frac{1}{N} \sum_{\substack{i: B^{(N)}_{im}=1,\\ C^{(N)}_{il}(t)=1}}\delta(x-x_i^{(N)}(t)) 
\end{equation}
and the proportion of these individuals compared to the total number of individuals in the system, $N$, by
$$  n^{(N)}_{m,l}(t) :=\int_D \rho_{m,l}^{(N)} (x,t) dx.$$  
The proportion $n^{(N)}_{m,l}(t)$ is constant in time for now, since influencer-follower relations cannot be changed.
Further, we denote the total empirical distribution of individuals that follow any influencer or medium by 
\begin{equation}\label{eq:total_emp} \rho^{(N)} (x,t) := \frac{1}{N} \sum_{i=1}^N\delta(x-x^{(N)}_i(t)) = \sum_{m=1}^M\sum_{l=1}^L \rho_{m,l}^{(N)} (x,t)
\end{equation}
with $\int_D \rho^{(N)} (x,t) dx = \sum_{m,l} n^{(N)}_{m,l}(t) =  1$.

The opinion dynamics of  individuals $i=1,\dots,N$ are given by \begin{equation}\label{eq:SDE}
dx^{(N)}_i(t) = F_i(\bm{x}^{(N)}, \bm{y}^{(N)}, \bm{z}^{(N)})  dt + \sigma dW_i(t).\end{equation} 
With the definition of the empirical distribution in~\eqref{eq:total_emp}, we can 
rewrite the interaction force $F_i$ by 
$$F_i(\bm{x}^{(N)}, \bm{y}^{(N)}, \bm{z}^{(N)}) = a  \frac{\int_D \rho^{(N)} (x,t) \phi(|x - x^{(N)}_i(t)|) (x-x^{(N)}_i(t))\, dx}{\int_D \rho^{(N)} (x,t) \phi(|x - x^{(N)}_i(t)|)\, dx}  + b\, (y_m^{(N)}(t)-x^{(N)}_i(t)) +c\, (z^{(N)}_l(t)-x^{(N)}_i(t)) $$
whenever the agent $i$ follows medium $m$ and influencer $l$. Thus the SDE in~\eqref{eq:SDE} depends no longer on the opinions of other individuals, but instead on the total empirical distribution $\rho^{(N)}$.

We are now interested in the limit $N\rightarrow \infty$. We assume that as $N$ grows, the number of individuals that follow a certain medium $m$ and a certain influencer $l$ grows like $ N n^{(N)}_{m,l}(t)$ with $n^{(N)}_{m,l}(t)$ independent of $N$.
It can then be expected \cite{sznitman1991topics,chaintron2021propagation} 
that as $N\rightarrow \infty$, any individual that follows a certain medium $m$ and influencer $l$ has an i.i.d. distributed opinion $\bar{x}_{m,l}$ that  satisfies the SDE 
\begin{equation}\label{eq:limit_opinion_process}
    d\bar{x}_{m,l}(t) = \mathcal{F}(\bar{x}_{m,l}, y_m, z_l,\rho)  dt + \sigma dW(t)
\end{equation}
with interaction force
$$\mathcal{F}(\bar{x}_{m,l}, y_m, z_l,\rho) = a  \frac{\int_D \rho (x,t) \phi(|x - \bar{x}_{m,l}|) (x-\bar{x}_{m,l})\, dx}{\int_D \rho (x,t) \phi(|x - \bar{x}_{m,l}|)\, dx}  + b\, (y_m(t)-\bar{x}_{m,l}) +c\, (z_l(t)-\bar{x}_{m,l}).$$
The opinion process $\bar{x}_{m,l}$ from~\eqref{eq:limit_opinion_process} depends on the density $\rho$  which can be interpreted as the limit of $\rho^{(N)}$ when $N\rightarrow \infty$. More precisely, denoting by $\mu_{m,l}(x,t)$ 
the probability density  of an individuals' opinion with medium~$m$ and influencer~$l$ at time $t$, $\rho_{m,l}(x,t) := n_{m,l}(t)\mu_{m,l}(x,t)$ denotes the probability density scaled by the proportion of these individuals in the system, and $\rho(x,t) = \sum_{m,l} \rho_{m,l}(x,t)$ denotes the probability density of any individual in the system. 
Stochastic processes of the form~\eqref{eq:limit_opinion_process} are also called \emph{McKean-Vlasov processes}.
Each scaled density $\rho_{m,l}$ fulfils a \emph{McKean Vlasov-type PDE}
\begin{eqnarray}\label{eq:mckean-pde}
\partial_t\rho_{m,l}(x,t) & = & \frac{1}{2} \sigma^2 \Delta \rho_{m,l}(x,t) - \nabla\cdot\left(\rho_{m,l}(x,t) \, \mathcal F(x, y_m, z_l,\rho) \right) 
\end{eqnarray} 
and we can also interpret each PDE for a fixed $m,l$ as describing the density of infinitely many copies of individuals that follow medium~$m$ and influencer~$l$. There is also a growing body of literature that considers an intermediate level of many but not infinitely many individuals, whose empirical distribution in the opinion space can approximately be described by a stochastic partial differential equation~\cite{dean1996langevin,kim2017stochastic,helfmann2021interacting,conrad2022feedback}.
 
\paragraph{Allowing influencer-follower connections to change.}
We now  additionally allow individuals to switch the influencer in time. In the ABM each individual $i$ can change the influencer to $l$ at the rate ${\Lambda_{m}^{\rightarrow l}}^{(N)}(x,t)$ where $m$ is the medium of individual $i$ and $x$ its current opinion. In the PDE we want these  changes from influencer $l'$ to $l$  to correspond to mass flowing from $\rho_{m,l'}$  to $\rho_{m,l}$. We will in the following derive the corresponding terms that have to be added to the PDE~\eqref{eq:mckean-pde} and are sometimes also called reaction terms. 

The total number of individuals that follow medium $m$ and influencer $l$ at time $t$ in the ABM is given by $ Y_{m,l}^{(N)}(t) := N \, n_{m,l}^{(N)}(t)$ and can be considered a jump process that only changes by $+1$ when an individual changes its influencer to $l$ or by $-1$ when an individual changes its influencer away from $l$.
The rate of \emph{any} individual changing from $(m,l')$ to $(m,l)$ is given by the sum of the individual change rates as follows
\begin{equation*}
    {\alpha_{m}^{l'\rightarrow l}}^{(N)}(t)  = \sum_{\substack{i:\;B^{(N)}_{im}=1,\\ C^{(N)}_{il'}(t)=1}}  {\Lambda_{m}^{\rightarrow l}}^{(N)}(x_i,t) = N \int_D {\Lambda_{m}^{\rightarrow l}}^{(N)}(x,t)\,  \rho_{m,l'}^{(N)}(x,t) dx.
\end{equation*}
With this, we can  write down the evolution of the jump process $Y_{m,l}^{(N)}(t)$ as
\begin{equation*}
    Y_{m,l}^{(N)}(t) = Y_{m,l}^{(N)}(0)+ \sum_{l' \neq l} \left( \mathcal{P}_{m}^{l'\rightarrow l}\left( \int_0^t {\alpha_{m}^{l'\rightarrow l}}^{(N)}(t') dt' \right)  - \mathcal{P}_{m}^{l\rightarrow l'}\left( \int_0^t {\alpha_{m}^{l\rightarrow l'}}^{(N)}(t')dt'\right)  \right)
\end{equation*} 
with unit-rate Poisson processes $\mathcal{P}_{m}^{l\rightarrow l'}$.

In the limit $N\rightarrow \infty$, the rates become large and we can replace the Poisson processes by their mean to get the \emph{reaction rate equation}~\cite{winkelmann2020stochastic} written in terms of the limiting proportions $n_{m,l}(t)$
\begin{equation}\label{eq:RRE}
\frac{dn_{m,l}}{dt}(t) = \sum_{l' \neq l} \left( n_{m,l'}(t) r_{m}^{l'\rightarrow l}(t) - n_{m,l}(t) r_{m}^{l\rightarrow l'}(t) \right) 
\end{equation}
with the spatially-averaged rates in the limit
\begin{equation*}r_{m}^{l'\rightarrow l}(t) = \frac{1}{n_{m,l'}(t)} \int_D \Lambda_{m}^{\rightarrow l}(x,t)\,  \rho_{m,l'}(x,t) dx\end{equation*}
and the limiting individual change rates
\begin{equation*}
\Lambda_{m}^{\rightarrow l}(x,t) =  \eta \, \psi(|z_l-x|) \, r\left(\frac{n_{m,l}(t)}{\sum_{m'} n_{m',l}(t)}\right).
\end{equation*}
The ODE~\eqref{eq:RRE} can be spatially extended to the following term~\cite{kim2017stochastic,helfmann2021interacting}
\begin{equation}\label{eq:reaction_term}
    \partial_t \rho_{m,l}(x,t) =  \sum_{l'\neq l} 
     - \Lambda_{m}^{\rightarrow l'}(x,t)\,  \rho_{m,l}(x,t) + \Lambda_{m}^{\rightarrow l}(x,t)\,  \rho_{m,l'}(x,t),
\end{equation} 
such that the complete PDE describing opinion changes and influencer changes is given by the sum of~\eqref{eq:mckean-pde} and~\eqref{eq:reaction_term}.

\subsection*{Limiting SDE dynamics of influencers and media}
The limiting dynamics of media and influencers follow by considering the limiting average position of followers of medium~$m$
\begin{eqnarray*}
    \tilde{x}^{(N)}_m(t) &=&  \frac{1}{\sum_{k=1}^N B^{(N)}_{km}} \sum_{i=1}^N B^{(N)}_{im} x^{(N)}_i(t)  \\
    &=& \frac{ \sum_{l=1}^L\int_D    x\, \rho^{(N)}_{m,l}(x,t) dx}{\sum_{l=1}^L n_{m,l}^{(N)}(t)}\\
    &\rightarrow &
    \frac{ \sum_{l=1}^L\int_D    x\, \rho_{m,l}(x,t) dx}{\sum_{l=1}^L n_{m,l}(t)} =: \tilde{x}_m(t)
\end{eqnarray*}
and of influencer~$l$
\begin{eqnarray*}
 \hat{x}^{(N)}_l(t) &=& \frac{1}{\sum_{k=1}^N C^{(N)}_{kl}(t)} \sum_{i=1}^N C^{(N)}_{il}(t) x^{(N)}_i(t) \\
 &=& \frac{ \sum_{m=1}^M\int_D    x\, \rho^{(N)}_{m,l}(x,t) dx}{\sum_{m=1}^M n^{(N)}_{m,l}(t)}  \\
 &\rightarrow& 
 \frac{ \sum_{m=1}^M\int_D    x\, \rho_{m,l}(x,t) dx}{\sum_{m=1}^M n_{m,l}(t)}   =:\hat{x}_{l}(t).
\end{eqnarray*}

\subsection*{No-flux boundary conditions}
We are using boundary conditions such that the total number of individuals in the system remains constant in time, i.e., we want $$\int_D  \partial_t\rho(x,t) dx = 0$$ for all $t$.  We can equivalently phrase this using the PDE equation and
the divergence theorem as
\begin{eqnarray*}
0 &=& \int_D  \partial_t\rho(x,t) dx\\
&=& \sum_{m,l} \int_D   \frac{1}{2} \sigma^2\Delta \rho_{m,l}(x,t) - \nabla\cdot \left(\rho_{m,l}(x,t) \mathcal F (x, y_m, z_l,\rho)\right)   dx \\
&=& \sum_{m,l}  \int_{dD}   \left( \frac{1}{2}\sigma^2 \nabla  \rho_{m,l}(x,t) -  \rho_{m,l}(x,t) \mathcal F (x, y_m, z_l,\rho)  \right) \cdot n \, dx
\end{eqnarray*}
where  $dD$ is the boundary of $D$, $n$ the unit outer normal to $D$. The terms for influencer changes disappeared since they only shift mass between the densities $\rho_{m,l}$ and therefore disappear when summing over $m,l$.
A sufficient condition to ensure mass conservation is therefore to ensure that the balance equation
$$ \frac{1}{2} \sigma^2\nabla  \rho_{m,l}(x,t) \cdot n =  \rho_{m,l}(x,t) \mathcal F (x, y_m, z_l,\rho)  \cdot n$$ holds everywhere on the boundary of $D$.

\bibliographystyle{plain}
\bibliography{main}

\end{document}